\documentclass[12pt,preprint]{aastex}
\begin{document}

\title{Interstellar Absorption Lines in the Direction of the Cataclysmic Variable SS~Cygni\footnote{Based on observations obtained with the Apache Point Observatory 3.5-m telescope, which is owned and operated by the Astrophysical Research Consortium, and on observations made with the NASA/ESA \emph{Hubble Space Telescope}, obtained from the Mikulski Archive for Space Telescopes. STScI is operated by the Association of Universities for Research in Astronomy, Inc., under NASA contract NAS5-26555.}}
\shorttitle{INTERSTELLAR LINES TOWARD SS CYG}
\shortauthors{RITCHEY ET AL.}

\author{Adam M. Ritchey and George Wallerstein}
\affil{Department of Astronomy, University of Washington, Seattle, WA 98195}
\email{aritchey@astro.washington.edu, wall@astro.washington.edu}
\and
\author{Jean McKeever}
\affil{Department of Astronomy, New Mexico State University, Las Cruces, NM 88001}
\email{jeanm12@nmsu.edu}

\begin{abstract}
We present an analysis of interstellar absorption lines in high-resolution optical echelle spectra of SS~Cyg obtained during an outburst in 2013 June and in archival \emph{Hubble Space Telescope} and \emph{Far Ultraviolet Spectroscopic Explorer} data. The Ca~{\sc ii}~K and Na~{\sc i}~D lines toward SS~Cyg are compared with those toward nearby B and A stars in an effort to place constraints on the distance to SS~Cyg. We find that the distance constraints are not very robust from this method due to the rather slow increase in neutral gas column density with distance and the scatter in the column densities from one sight line to another. However, the optical absorption-line measurements allow us to derive a precise estimate for the line-of-sight reddening of $E(B-V)=0.020\pm0.005$ mag. Furthermore, our analysis of the absorption lines of O~{\sc i}, Si~{\sc ii}, P~{\sc ii}, and Fe~{\sc ii} seen in the UV spectra yields an estimate of the H~{\sc i} column density and depletion strength in this direction.
\end{abstract}

\section{INTRODUCTION}
SS~Cygni was first recognized to be a variable star from observations by Ms. Louisa Wells \citep{pic96}. Its light curve was first published by \citet{par00}, who found that it suffered outbursts from its usual brightness of 11.3 mag to 8.5 mag (in the photographic band then in use) at irregular intervals of about 50 days. As one of the nearest members, and indeed a prototype, of the class of variable stars known as cataclysmic variables, an accurate determination of the distance and reddening along the line of sight to SS~Cyg is vital to understanding the physics of the outbursts of these stars. Recently, \citet{mil13} derived a distance of $114\pm2$~pc for SS~Cyg using very long baseline interferometric radio observations. Their distance disagrees considerably with previous distance estimates obtained using the Fine Guidance Sensors (FGS) of the \emph{Hubble Space Telescope} (\emph{HST}) by \citet{har99,har00,har04}. The most recent paper by this group reports a distance of $165\pm12$~pc for SS~Cyg \citep{har04}.

Fifty years ago, one of us endeavored to compare the strengths of interstellar absorption lines toward SS~Cyg with those of stars in the same direction in order to constrain the absolute magnitude and foreground reddening of the cataclysmic variable \citep{wal63a}. No interstellar Ca~{\sc ii} lines were seen in the direction of SS~Cyg on two photographic plates obtained with the coude spectrograph of the 120-inch telescope of the Lick Observatory during two outbursts in 1960. With the availability of CCD detectors and access to an echelle spectrograph, and in light of the recent controversy regarding the distance to SS~Cyg, it seemed worthwhile to repeat that experiment. After completing our analysis, however, we became aware of the paper by \citet{nel13}, who reanalyzed the \emph{HST}/FGS data for SS~Cyg and found a distance of $120\pm6$~pc, statistically in agreement with the distance measurement of \citet{mil13}. Since this latest result by \citet{nel13} seems to have settled the controversy regarding the distance to SS~Cyg, and since the interstellar absorption method cannot, in any case, be definitive, we turned our attention to other aspects of the line of sight in this direction.

Both the interstellar reddening and H~{\sc i} column density toward SS~Cyg also seem to be mildly controversial. \citet{mau88} determined a hydrogen column density of log~$N$(H~{\sc i})~=~19.55 based on the assumption that the observed lines of N~{\sc i} and S~{\sc ii} represented species that were undepleted with respect to solar abundances. From this result, they estimated the reddening along the line of sight to be $E(B-V)=0.007$ mag using the correlation between $N$(H~{\sc i}) and $E(B-V)$ obtained by \citet{boh78} for ``intercloud'' stars. This estimate for the reddening was somewhat at odds with an earlier determination by \citet{ver87} of $E(B-V)=0.04\pm0.03$ mag based on the strength of the 2200~\AA{} extinction bump (although the two estimates are nominally consistent within the large relative uncertainties). Moreover, \citet{sah06} had to adopt an H~{\sc i} column density of log~$N$(H~{\sc i})~=~20.0 in order to model the Lyman series lines observed in their \emph{Far Ultraviolet Spectroscopic Explorer} (\emph{FUSE}) data of SS~Cyg. More recently, \citet{gau11} found that the reddening toward SS~Cyg is comprised of two components, a constant interstellar component and a variable component, which is intrinsic to the binary system.

In this paper, we present new optical observations of SS~Cyg obtained during a recent outburst and examine the interstellar absorption lines in this direction and in other nearby directions. We also take the opportunity to analyze archival UV spectra of SS~Cyg that were acquired with \emph{HST} and \emph{FUSE}. Together, the optical and UV data allow us to derive new estimates for the interstellar reddening and H~{\sc i} column density in the direction of this prototypical cataclysmic variable.

\section{OBSERVATIONS AND PROFILE FITTING}
\subsection{Apache Point Observatory Data}
On 2013 June 13, we obtained two 10 minute exposures of SS~Cyg with the echelle spectrograph of the 3.5-m telescope at Apache Point Observatory (APO). The echelle provides complete wavelength coverage in the range 3800--10200 \AA{} at a resolving power of $R=31,500$. The raw data were reduced using standard procedures within the IRAF environment. The individual extracted spectra were co-added to yield a signal-to-noise ratio (S/N) of about 160 in the continuum near the Na~{\sc i}~D lines and 90 near the Ca~{\sc ii}~K line. Very weak, broad emission was noted in the Ca~{\sc ii}~H and K lines at the radial velocity of SS~Cyg, though narrow interstellar features were also easily discernible. Telluric absorption lines near the interstellar Na~{\sc i}~D features were removed through division by an unreddened, early-type star. The K~{\sc i} line at 7698 \AA{} was not seen, though the S/N was about 200 and the largest noise dip indicated an upper limit to the equivalent width of 3 m\AA. Neither CH nor CH$^+$ is present in the co-added spectrum.

On 2013 July 4, we observed 8 stars of type B8 to A2 within approximately 5$^{\circ}$ of SS~Cyg so that we could compare the strengths of the interstellar lines in this region. The data were reduced following the same procedures as for SS~Cyg. Unfortunately, two of the comparison stars (HD~206280 and HR~8349) exhibit narrow stellar lines making them unsuitable for interstellar measurements and so were removed from the analysis. Basic data for the remaining comparison stars are shown in Table~1. The quoted distances are from \emph{Hipparcos} measurements of at least 8$\sigma$ significance. Equivalent widths and column densities for the interstellar absorption lines of Ca~{\sc ii}~K, Na~{\sc i}~D$_2$ and D$_1$, and K~{\sc i}~$\lambda7698$ were determined for each star through multi-component Voigt profile fitting of the observed spectra. The analysis was relatively straightforward since at most there were two components along a given line of sight. The resulting total equivalent widths and column densities are presented in Table~2. Figure~1 provides an example of the APO data and the fitting procedure, showing fits for the Ca~{\sc ii}~K and Na~{\sc i}~D lines toward SS~Cyg.

\subsection{Archival \emph{HST}/STIS Spectra}
High-resolution UV spectra of SS~Cyg acquired with the Space Telescope Imaging Spectrograph (STIS) of \emph{HST} were obtained from the Mikulski Archive for Space Telescopes (MAST). Only those spectra acquired using the E140H grating (with the central wavelength set to 1598 \AA) were retrieved from the archive. For these spectra, the $0.2\times0.2$ arcsec slit was used yielding a resolving power of $R=82,000$. The data cover the wavelength region from 1495 \AA{} to 1690 \AA{}. All of the E140H exposures, obtained on four separate visits during an outburst in 1999 June and totaling 16,000 seconds of exposure time, were co-added to increase the S/N in the final spectrum. Prominent interstellar lines of C~{\sc i}, C~{\sc iv}, Si~{\sc ii}, Fe~{\sc ii}, and Al~{\sc ii} are clearly present\footnote{The interstellar C~{\sc iv} absorption features have been interpreted by \citet{mau88} as probing an expanding H~{\sc ii} region photoionized by the EUV and soft X-ray flux of SS~Cyg. In the STIS spectra examined here, the C~{\sc iv} lines show a much broader velocity distribution and are displaced significantly toward more negative velocities compared to the other interstellar features.}. We also detect a weak feature associated with P~{\sc ii}~$\lambda1532$ in the co-added spectrum. These data were originally discussed in \citet{mau04}, but no detailed description was given of the interstellar features present.

Here, we focus our analysis on lines from atoms in their dominant state of ionization in neutral, diffuse clouds. That is, we derive column densities for Si~{\sc ii}~$\lambda1526$, P~{\sc ii}~$\lambda1532$, Fe~{\sc ii}~$\lambda1608$, and Al~{\sc ii}~$\lambda1670$. As for the optical data, column densities were derived through Voigt profile fits of the observed spectra. For each fit, we assume two components are present along the line of sight, but the velocity separation between the components is allowed to vary from one species to the next. The total equivalent widths and column densities resulting from these fits are given in Table~3, while the fits themselves are presented in Figure~2.

\subsection{Archival \emph{FUSE} Data}
In order to perform the analysis described in Section~4, a reliable measurement of the O~{\sc i} column density toward SS~Cyg is required. For this, we obtained archival \emph{FUSE} spectra from the MAST archive. \citet{sah06} reported an O~{\sc i} column density of log~$N$(O~{\sc i})~=~17.0 from \emph{FUSE} observations of SS~Cyg acquired during an outburst in 2000 November, but they caution that their determination was based primarily on saturated lines (using a curve-of-growth analysis) and likely suffers from large uncertainties. We examine the same observations that they used but focus on weaker O~{\sc i} lines and derive our column density through profile fitting. Following \citet{sah06}, we consider only the observations on 2000 November 2, when the source was the brightest and where the S/N is the highest. The 13 individual exposures for each channel and detector segment were cross-correlated to place them on the same velocity scale and then co-added. The O~{\sc i} lines considered here, O~{\sc i}~$\lambda925$ and O~{\sc i}~$\lambda930$, reside in the SiC-2A and SiC-1B segments. Portions of the spectra covering these lines were extracted and shifted according to the velocity of the Fe~{\sc ii}~$\lambda1608$ feature seen in STIS data, and then the portions covering the same wavelengths from the two segments were combined. The nominal resolving power of \emph{FUSE} is $R\sim20,000$.

Figure~3 presents our profile fits to the two O~{\sc i} lines from \emph{FUSE}, where we again assume that there are two velocity components along the line of sight. The total equivalent widths and column densities are given in Table~3. The third line in Table~3 gives the weighted mean of the column densities from the two O~{\sc i} transitions and the corresponding abundance and depletion values. We note that our determination of log~$N$(O~{\sc i}) is considerably smaller than that of \citet{sah06}, but we also note that in Figure~1 of that paper their model spectrum clearly exceeds the absorption profile for the O~{\sc i} line at 930 \AA{}, indicating that they may have overestimated the column density.

\section{THE DISTANCE AND REDDENING TO SS CYGNI FROM OPTICAL ABSORPTION LINES}
The initial purpose of this project was to place constraints on the distance to SS~Cyg by comparing the strength of the interstellar features in its direction with those of nearby directions. Ultimately, we were unable to place robust constraints on the distance using this method (for reasons discussed below), but we do uncover some interesting properties of the interstellar medium (ISM) in this direction. In Figure~4, we plot the column densities of Na~{\sc i} and Ca~{\sc ii} derived from our APO observations (solid symbols) as a function of the distance to the background star. To supplement our sample and to extend our analysis to more distant stars, we include the Na~{\sc i} and Ca~{\sc ii} measurements compiled in \citet{wel10} toward stars in the same part of the sky as those we observed. As with our sample, we only include stars with \emph{Hipparcos} distances that are 8$\sigma$ or better. These measurements are shown as open symbols in Figure~4. The $\times$'s in Figure~4 denote the extremes of the more recent distance estimates reported for SS~Cyg, where the lower value corresponds to the radio interferometric result of $114\pm2$~pc from \citet{mil13} and the upper value to the prior \emph{HST} determination of $165\pm12$~pc from \citet{har04}.

First, we note that there is a steep drop in the column density of neutral gas for distances less than about 85 pc, consistent with the much more extensive analysis of \citet{wel10}. This corresponds to the wall of neutral material that forms the outer boundary of the Local Bubble, a cavity of rarefied gas encompassing the Sun. The boundary wall is most clearly seen in the plot of Na~{\sc i} column densities, but is also indicated in the Ca~{\sc ii} data in the sense that 2 out of the 3 sight lines with path lengths less than 85 pc have no detectable Ca~{\sc ii}. The stars HR 8489 and 79 Cyg, which nominally have almost identical distances ($82\pm2$ and $84\pm2$ pc, respectively), are interesting because they show quite different column densities of Na~{\sc i} and Ca~{\sc ii}. Evidently, HR 8489 lies just beyond and 79 Cyg just in front of the boundary wall in their respective directions, which are separated by approximately 8$^{\circ}$ on the sky.

For distances larger than 85 pc, the column density of neutral material shows a more steady increase. This increase amounts to 0.0015 dex pc$^{-1}$ for Na~{\sc i} and 0.0011 dex pc$^{-1}$ for Ca~{\sc ii}, based on least-squares fits to the column densities for path lengths greater than 85 pc (see Figure~4). Thus, while the slopes for Na~{\sc i} and Ca~{\sc ii} are nearly the same, the scatter in the measurements is quite different. If we focus only on those sight lines between 100 and 200 pc in length, the scatter in the Na~{\sc i} column densities is 0.33 dex, while for the same sight lines, the scatter in the Ca~{\sc ii} column densities is only 0.08 dex. (This result for Ca~{\sc ii} excludes the outlier at 150 pc, which is also excluded in the above result for Na~{\sc i} because Na~{\sc i} is not detected in this direction.) The difference in dispersion between the Na~{\sc i} and Ca~{\sc ii} column densities is again consistent with the results presented in \citet{wel10}. The interpretation is that Na~{\sc i} probes denser clumps of gas confined to smaller volumes, while Ca~{\sc ii} is more broadly distributed throughout the outer cloud envelopes. Thus, whether or not a given sight line passes through a dense clump will affect the resulting Na~{\sc i} column density more strongly than that of Ca~{\sc ii}\footnote{We also note that the scatter in $N$(Na~{\sc i}) is mirrored in the column densities of K~{\sc i} listed in Table~2.}.

Taken together, the rather shallow increase in the column density of neutral gas as a function of distance and the dispersion in the measurements from one sight line to another preclude a robust determination of the distance to SS~Cyg based on the strength of the interstellar absorption lines. The difference in the column density that one would expect if SS~Cyg were positioned at 165 pc rather than 114 pc is only about 0.07 dex, smaller even than the dispersion in Ca~{\sc ii} column densities. Still, we can say with high confidence that SS~Cyg lies beyond the outer boundary of the Local Bubble. Of course, this point may now be moot since the recent reanalysis of the \emph{HST}/FGS data by \citet{nel13} essentially confirms the distance estimate of \citet{mil13} of 114 pc.

While our optical observations of SS~Cyg may not significantly constrain the distance to the cataclysmic variable, they can help to establish the reddening in this direction due to the well-known correlation between Na~{\sc i}~D absorption and dust extinction. This relationship was most recently quantified by \citet{poz12}. From our measured equivalent widths of $W_{\lambda}$(D$_2$)~=~$78.8\pm1.3$~m\AA{} and $W_{\lambda}$(D$_1$)~=~$62.7\pm1.3$~m\AA{} and using the relations in \citet{poz12}, we find reddening values of $E(B-V)=0.018\pm0.007$ mag from D$_2$ and $E(B-V)=0.025\pm0.010$ mag from D$_1$, where the errors include both the errors in the equivalent width measurements and the scatter in the relations. The weighted mean of the results from the two Na~{\sc i}~D lines is then $E(B-V)=0.020\pm0.005$ mag. This result is intermediate between those of \citet{mau88}, who found $E(B-V)=0.007$ mag, and \citet{ver87}, who give $E(B-V)=0.04\pm0.03$ mag. The latter value is identical to the interstellar value reported by \citet{gau11} of $E(B-V)=0.04\pm0.02$ mag, which can be said to be in nominal agreement with our result. Applying the \citet{poz12} relations to the other stars observed in the direction of SS~Cyg results in $E(B-V)$ values in the range 0.01--0.04 mag, similar to the values listed for these stars in Table~1.

\section{AN ESTIMATE OF THE HYDROGEN COLUMN DENSITY FROM UV ABSORPTION LINES}
The high-resolution UV spectra of SS~Cyg obtained from the \emph{HST} and \emph{FUSE} archives allow us to examine the abundances and depletions along this line of sight through the local ISM. However, before an evaluation can be made of the elemental abundances, knowledge of the total hydrogen column density is required\footnote{Along lines of sight to cataclysmic variables, in particular, the hydrogen column density is also an important parameter for constraining the soft X-ray luminosity of the variable \citep[See][]{mau88}.}. As mentioned in the introduction, the presently available estimates of the hydrogen column density toward SS~Cyg appear to be in conflict, with the value reported by \citet{sah06} being about a factor of 3 larger than that determined by \citet{mau88}. Since \citet{sah06} based their result on fits to the H~{\sc i} Lyman series lines in their \emph{FUSE} spectra, while \citet{mau88} considered the abundances of undepleted elements, part of the discrepancy could be related to an uncertain contribution to the H~{\sc i} absorption from circumstellar gas in the accretion disk of SS~Cyg. Indeed, the H~{\sc i} Lyman $\alpha$ line seen in low-resolution \emph{HST} spectra available from the MAST archive appears to be significantly blue-shifted away from the expected interstellar velocity in accordance with the radial velocity of SS~Cyg itself.

In order to avoid the issue of potential contamination of the H~{\sc i} lines by circumstellar material, we use an indirect method of estimating the total hydrogen column density, which was formulated by \citet{jen09} and used successfully to derive synthetic values of $N$(H~{\sc i}) for white dwarf stars in the Local Bubble. The method is based on the fact that the depletions of different elements follow distinct but predictable patterns according to the overall amount of depletion present along a given line of sight. Thus, if the relative abundances for a set of elements with different depletion characteristics can be determined, then it is possible to estimate both the overall depletion strength and the line-of-sight hydrogen column density. In practical terms, the method amounts to performing a least-squares linear fit using the expressions defined in Equations (24a)--(24e) of \citet{jen09} and we refer the reader to that paper for more details.

In order for the method to be effective, column densities for elements with a range of depletion properties must be determined. The crucial parameter, in particular, is the rate at which the element depletes (i.e., the depletion slope). Since the \emph{HST}/STIS data allowed us to derive column densities for moderately-depleting to more heavily-depleting elements like P, Si, and Fe, we required a column density for a more lightly-depleting element like O. This was the purpose of obtaining the \emph{FUSE} spectra as described in Section~2.3. Figure~5 presents the linear fit to the quantities $y$ and $x$ as defined in Equations (24b) and (24c), respectively, of \citet{jen09}, where the column densities for O~{\sc i}, Si~{\sc ii}, P~{\sc ii}, and Fe~{\sc ii} come from Table~3. Our measurement of the Al~{\sc ii} column density was not used since this element was not considered by \citet{jen09}. The slope of the linear fit represents the depletion strength for the line of sight (denoted $F_*$ in Jenkins 2009)\footnote{The depletion strength $F_*$ is a dimensionless quantity that equals 0.0 for sight lines showing almost no depletion and 1.0 for sight lines through typical cold, diffuse clouds like that toward $\zeta$~Oph.}, while the $y$-intercept represents the logarithm of the hydrogen column density. For the fit presented in Figure~5, we find $F_*=0.15\pm0.26$ and log~$N$(H)~=~$19.49\pm0.26$. This result for log~$N$(H) could also be labeled log~$N$(H~{\sc i}) because there is very little H$_2$ along the line of sight \citep{sah06}. Our result for log~$N$(H~{\sc i}) is thus in very good agreement with that of \citet{mau88}. Furthermore, the value of $F_*$ that we derive is very similar to values derived by \citet{jen09} for white dwarf stars in the Local Bubble.

\section{CONCLUDING REMARKS}
We examined new optical echelle observations and reanalyzed high-resolution archival UV spectra of the prototypical cataclysmic variable SS~Cyg in order to investigate various properties of the line of sight in this direction. By comparing the Na~{\sc i} and Ca~{\sc ii} column densities toward SS~Cyg with those of nearby stars, we attempted to place limits on the distance to the cataclysmic variable, which could be compared with other more precise estimates. We find that the distance constraints from this method are not very robust due to the very gradual increase in neutral column density with distance (for stars beyond 85~pc) and the scatter in the column densities from one sight line to another. Still, the optical absorption-line measurements provide a precise value for the interstellar reddening toward SS~Cyg, which is free of any contamination by a reddening source intrinsic to the binary system, as discussed in \citet{gau11}. Finally, we used the column densities of dominant ions obtained from archival \emph{HST} and \emph{FUSE} spectra to estimate the depletion strength and the total hydrogen column density in the direction of SS~Cyg. Our estimate for log~$N$(H~{\sc i}) is in very good agreement with that of \citet{mau88}, indicating that the direct modelling of the Lyman series lines in \emph{FUSE} spectra by \citet{sah06} may be affected by a contribution to the H~{\sc i} absorption from the accretion disk of SS~Cyg.

\acknowledgements
Support for this research was provided by the Kennilworth Fund of the New York Community Trust. We thank Myra Stone for her help in preparing this paper for publication. The results presented here are based on observations obtained with the Apache Point Observatory 3.5-m telescope, which is owned and operated by the Astrophysical Research Consortium, and on observations made with the NASA/ESA \emph{Hubble Space Telescope}, obtained from the Mikulski Archive for Space Telescopes. STScI is operated by the Association of Universities for Research in Astronomy, Inc., under NASA contract NAS5-26555.

\bibliography{ms}

\begin{deluxetable}{lcccccccc}
\tablenum{1}
\tablecolumns{9}
\tablewidth{0pc}
\tabletypesize{\footnotesize}
\tablecaption{Comparison Stars in the Direction of SS Cygni\label{t1}}
\tablehead{
\colhead{Star}  &  \colhead{Sp. Type}  &  \colhead{$d$}  & \colhead{$l$}  & \colhead{$b$}  &\colhead{$V$}  &  \colhead{$B-V$}  &  \colhead{$E(B-V)$}  &  \colhead{$v$~sin~$i$} \\
\colhead{} & \colhead{} & \colhead{(pc)} & \colhead{(deg)} & \colhead{(deg)} & \colhead{(mag)} & \colhead{(mag)} & \colhead{(mag)} & \colhead{(km s$^{-1}$)} \\
}
\startdata
74 Cyg  &  A5V  &  $66\pm2$  &  87.61  &  \phn$-$8.76   &  5.05  &  \phantom{+}0.18  &  \phantom{+}0.03  &  171  \\
HR 8489 &  A2V  &  $82\pm2$  &  94.81  &  $-$11.20  & 5.72  &  \phantom{+}0.00  &  $-$0.05  &  245  \\ 
79 Cyg  &  A0V  &  $84\pm2$  &  87.11  &  $-$11.18  & 5.70  &  \phantom{+}0.00  &  \phantom{+}0.02  &  $<17$  \\ 
77 Cyg\tablenotemark{a}  &  A0V  & $126\pm6\phn$  &  88.84  &  \phn$-$8.96  &  5.73  &  \phantom{+}0.05  &  \phantom{+}0.07  &  45  \\
HR 8338 &  B8V  &  $146\pm7\phn$ &  88.12  &  $-$11.56  &  6.09  & $-$0.07  &  \phantom{+}0.04  &  100  \\
76 Cyg  &  A2V  &  $152\pm8\phn$ &  88.54  &  \phn$-$9.06  &  6.08  &  \phantom{+}0.06  &  \phantom{+}0.01  &  150  \\
\enddata
\tablenotetext{a}{A double-line spectroscopic binary.}
\end{deluxetable}

\begin{deluxetable}{lccccccc}
\tablenum{2}
\tablecolumns{8}
\tablewidth{0pc}
\tabletypesize{\footnotesize}
\tablecaption{Total Equivalent Widths and Column Densities for Optical Lines\label{t2}}
\tablehead{
\colhead{Star}  &  \colhead{$W_{\lambda}$(K)}  &  \colhead{log $N$(Ca~{\sc ii})}  &  \colhead{$W_{\lambda}$(D$_2$)}  &  \colhead{$W_{\lambda}$(D$_1$)}  &  \colhead{log $N$(Na~{\sc i})\tablenotemark{a}}  &  \colhead{$W_{\lambda}$(7698)}  &  \colhead{log $N$(K~{\sc i})}  \\
\colhead{}  &  \colhead{(m\AA)}  &  \colhead{}  &  \colhead{(m\AA)}  &  \colhead{(m\AA)}  &  \colhead{}  &  \colhead{(m\AA)}  &  \colhead{}  \\
}
\startdata
SS Cyg  &  $20.1\pm1.8$  &  $11.40\pm0.04$  &  \phn$78.8\pm1.3$  &  \phn$62.7\pm1.3$  &  $12.05\pm0.03$  &  $<3$  &  $<10.3$  \\
74 Cyg  &  $<11$  &  $<11.1$  &  \phn$10.8\pm1.2$  &  \phn\phn$6.1\pm1.2$  &  $10.76\pm0.04$  &  $<3$  &  $<10.2$  \\
HR 8489  &  $23.9\pm2.1$  &  $11.63\pm0.04$  &  $127.5\pm1.0$  &  $107.6\pm1.0$  &  $12.45\pm0.03$  &  $10.0\pm0.9$  &  $10.78\pm0.04$  \\
79 Cyg  &  $<11$  &  $<11.1$  &  \phn$37.1\pm0.9$  &  \phn$20.0\pm0.9$  &  $11.39\pm0.01$  &  $<3$  &  $<10.2$  \\
77 Cyg  &  $32.1\pm5.3$  &  $11.61\pm0.07$  &  $181.9\pm1.4$  &  $130.1\pm1.4$  &  $12.48\pm0.02$  &  $13.2\pm1.1$  &  $10.90\pm0.03$  \\
HR  8338  &  $34.6\pm3.2$  &  $11.64\pm0.04$  &  $128.9\pm1.4$  &  \phn$83.5\pm1.4$  &  $11.94\pm0.01$  & \phn$8.0\pm1.4$  &  $10.67\pm0.07$  \\
76 Cyg  &  $36.7\pm3.3$  &  $11.72\pm0.04$  &  $219.5\pm1.4$  &  $172.5\pm1.3$  &  $12.64\pm0.10$  &  $22.6\pm1.0$  &  $11.15\pm0.02$  \\
\enddata
\tablenotetext{a}{Weighted mean of the column densities derived from the Na~{\sc i}~D$_2$ and D$_1$ lines.}
\end{deluxetable}

\begin{deluxetable}{lccccc}
\tablenum{3}
\tablecolumns{6}
\tablewidth{0pc}
\tabletypesize{\footnotesize}
\tablecaption{Abundances and Depletions from UV Lines toward SS Cyg\label{t3}}
\tablehead{
\colhead{Species}  &  \colhead{$\lambda$}  &  \colhead{$W_{\lambda}$}  &  \colhead{log $N$(X)}  &  \colhead{log (X/H)\tablenotemark{a}}  &  \colhead{[X/H]\tablenotemark{b}}  \\
\colhead{}  &  \colhead{(\AA)}  &  \colhead{(m\AA)}  &  \colhead{}  &  \colhead{}  &  \colhead{}  \\
}
\startdata
\multicolumn{6}{c}{\emph{FUSE}} \\
\hline
O~{\sc i}  &  \phn925.446  &  $26.6\pm2.0$  &  $16.16\pm0.04$  &  \ldots  & \ldots  \\
           &  \phn930.257  &  $33.0\pm2.1$  &  $16.13\pm0.04$  &  \ldots  & \ldots  \\
           &  \ldots  &  \ldots  &  $16.15\pm0.03$\tablenotemark{c}  &  $-3.34\pm0.26$  &  $-0.10\pm0.27$  \\
\hline
\multicolumn{6}{c}{STIS} \\
\hline
Si~{\sc ii}  &  1526.707  &  $92.2\pm0.8$  &  $14.36\pm0.03$  &  $-5.13\pm0.26$  &  $-0.74\pm0.26$  \\
P~{\sc ii}  &  1532.533  &  \phn$2.4\pm0.7$  &  $13.60\pm0.11$  &  $-5.89\pm0.28$  &  $+0.57\pm0.29$  \\
Fe~{\sc ii}  &  1608.451  &  $51.7\pm1.0$  &  $13.99\pm0.04$  &  $-5.50\pm0.26$  &  $-1.04\pm0.27$  \\
Al~{\sc ii}  &  1670.789  &  $86.4\pm1.2$  &  $12.82\pm0.03$  &  $-6.67\pm0.26$  &  $-1.21\pm0.26$  \\
\enddata
\tablenotetext{a}{Elemental abundance, defined as log~$N$(X)~$-$~log~$N$(H), adopting a total hydrogen column density of log~$N$(H)~=~$19.49\pm0.26$ (see Section~4).}
\tablenotetext{b}{Depletion, defined as log~(X/H)~$-$~log~(X/H)$_{\odot}$, using solar abundances from \citet{lod03}.}
\tablenotetext{c}{Weighted mean of the column densities from the two O~{\sc i} transitions.}
\end{deluxetable}

\clearpage

\begin{figure}
\centering
\includegraphics[width=0.45\textwidth]{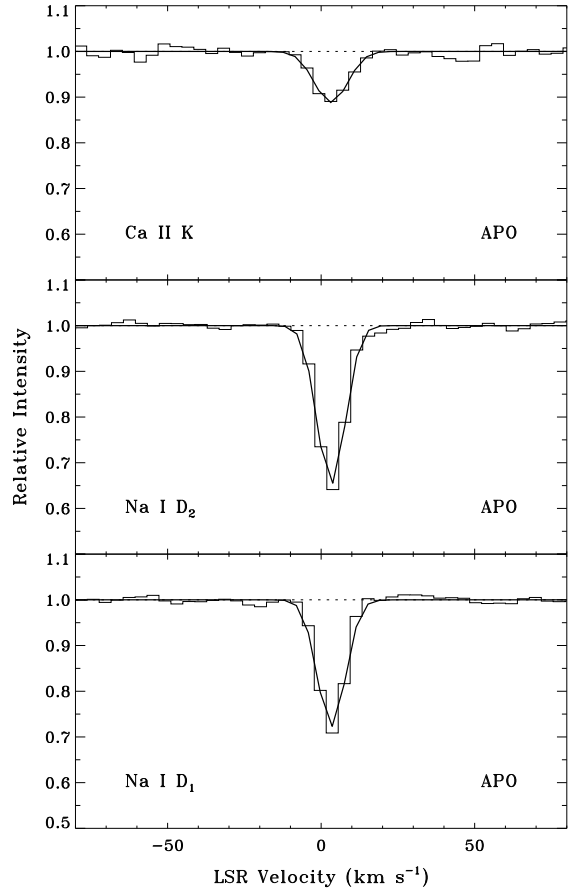}
\caption{Profile synthesis fits to the Ca~{\sc ii}~K and Na~{\sc i}~D lines toward SS~Cyg. The histogram indicates the observed spectrum from APO, while the smooth curve represents the synthetic profile.}
\label{f1}
\end{figure}

\begin{figure}
\centering
\includegraphics[width=0.45\textwidth]{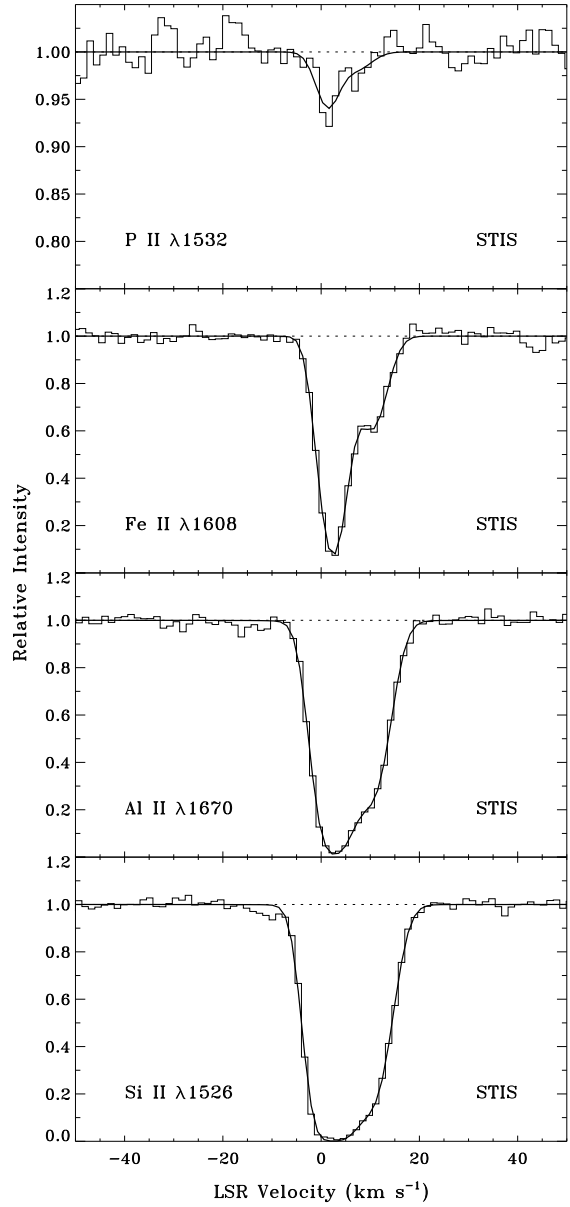}
\caption{Profile synthesis fits to the P~{\sc ii}, Fe~{\sc ii}, Al~{\sc ii}, and Si~{\sc ii} lines toward SS~Cyg. The histogram indicates the observed spectrum obtained with STIS, while the smooth curve represents the synthetic profile.}
\label{f2}
\end{figure}

\begin{figure}
\centering
\includegraphics[width=0.45\textwidth]{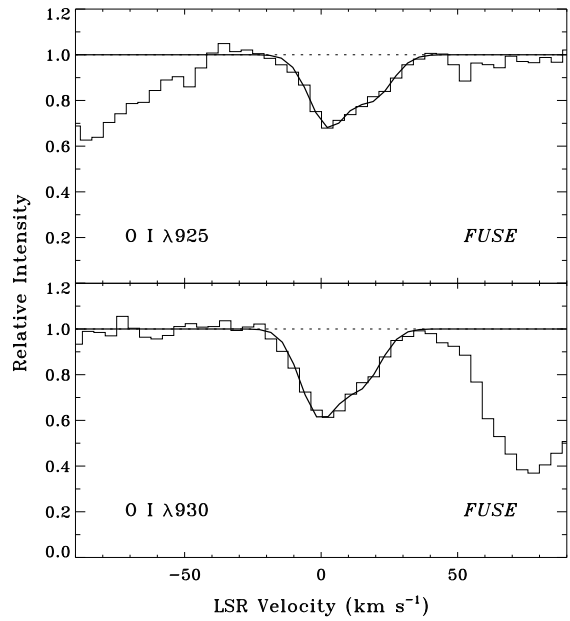}
\caption{Profile synthesis fits to the O~{\sc i}~$\lambda925$ and O~{\sc i}~$\lambda930$ lines toward SS~Cyg. The histogram indicates the observed spectrum obtained with \emph{FUSE}, while the smooth curve represents the synthetic profile. The additional strong absorption features are due to interstellar H$_2$.}
\label{f3}
\end{figure}

\begin{figure}
\centering
\includegraphics[width=0.6\textwidth]{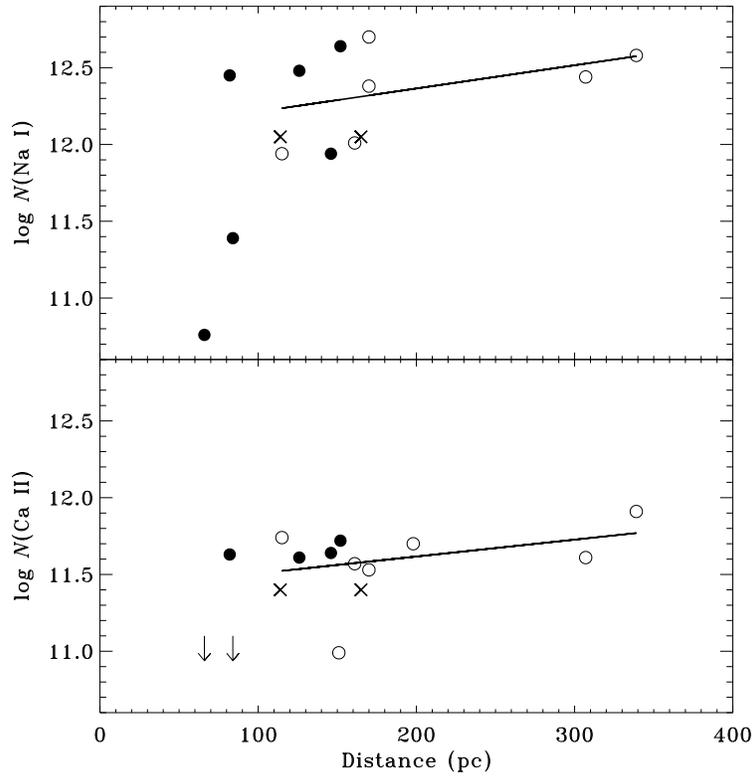}
\caption{Column densities of Na~{\sc i} (\emph{upper}) and Ca~{\sc ii} (\emph{lower}) as a function of distance for stars in the vicinity of SS~Cyg. Solid symbols (and upper limits) are our measurements. Open symbols are from \citet{wel10}. A least-squares linear fit is shown for sight lines to stars beyond 85 pc. The two $\times$'s in each panel denote the range of distance estimates for SS~Cyg.}
\label{f4}
\end{figure}

\begin{figure}
\centering
\includegraphics[width=0.6\textwidth]{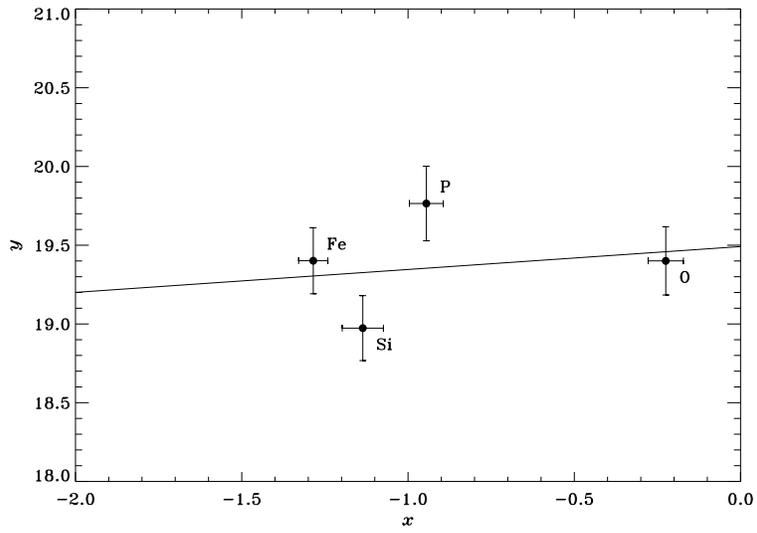}
\caption{An application of the methodology described in \citet{jen09} to estimate the total hydrogen column density and the depletion strength for the line of sight to SS~Cyg.}
\label{f5}
\end{figure}

\end{document}